\documentclass[aps,prb,superscriptaddress,twocolumn,showpacs]{revtex4-1}
\usepackage{amsfonts,amsmath,amssymb,bm,esvect,verbatim}
\usepackage{blindtext,chngcntr,color,colortbl,epsfig,graphicx,mathtools,xr-hyper,hyperref,multirow,tabularx,tikz}
\allowdisplaybreaks

\newcommand{\nc}{\newcommand}

\nc{\nn}{\nonumber \\}
\nc{\ds}{\displaystyle}

\DeclarePairedDelimiter\ket{\lvert}{\rangle}
\DeclarePairedDelimiterX\braket[2]{\langle}{\rangle}{#1 \delimsize\vert #2}

\nc{\commentout}[1]{\marginpar{#1}}

\graphicspath{{../figures/}{./figures/}{../../figures/}} 

\begin{document}

\title{Measurement-based Uncomputation Applied to Controlled Modular Multiplication}

\author{Panjin Kim}
\affiliation{The Affiliated Institute of ETRI, Daejeon 34044, Korea}
\email{pansics@nsr.re.kr}
\author{Daewan Han}
\affiliation{The Affiliated Institute of ETRI, Daejeon 34044, Korea}

\begin{abstract}
This is a brief report on a particular use of measurement-based uncomputation. Though not appealing in performance, it may shed light on optimization techniques in various quantum circuits.
\end{abstract}

\maketitle

Circuit implementations of quantum algorithms often confront a need for uncomputation to get rid of intermediate information.
As an example relevant to this work, all known circuit implementations of Shor algorithm~\cite{shor94} carefully deal with the issue.
See, Beauregard's comment on a garbage bit appearing in his modular addition~\cite{beauregard03}.
The usual way of handling the uncomputation is to apply Bennett's method~\cite{bennett73}, but recent progress in the field explores a new direction~\cite{gidney18}.

To illustrate the idea of measurement-based uncomputation, consider a bijective function $f: x \mapsto y$ and its standard quantum implementation $U_f:\ket{x}\ket{0} \mapsto \ket{x}\ket{y}$.
Since $f$ is bijective, the information on $x$ can be completely removed while $y$ remains still~\cite{kitaev97}.
An immediate way to do so is to find a circuit for $f^{-1}$ and applying it, typically resulting in doubling the cost.
Now instead of computing $f^{-1}$, assume a unitary transformation is applied to the first register locally, leading to a state $\sum_{x'} c(x',y) \ket{x'} \ket{y}$, where $c(x',y)$ is a probability amplitude.
The probability amplitude is dependent not only on $x'$ but also on $y$ if $x$ and $y$ have been entangled in the first place.
If one carries out the measurement on the first register getting $X'$, the state would collapse into $\big[ c(X',y) / \left|c(X',y) \right| \big] \ket{X'}\ket{y}$.
The phase $c(X',y)/|c(X',y)|$ can be corrected by manipulating the second register since $X'$ is known and the register is holding the information on $y$.
The first register is cleanly separated from the second register, thus $X'$ can safely be returned to 0 by any local means.

One emphasis must be put on the efficiency of the method.
Computation-based uncomputation allows more concrete resource estimates for the algorithms as the number of gates and qubits would be the only concern, whereas measurement-based one further requires intermediate measurements and classical feedback processes which can hardly be compared with other quantum resources as of the time of writing.
Readers are kindly advised to take this direction as one of the options even if it reduces the gate depth without involving more qubits.

This report examines the use of measurement-based uncomputation in controlled modular multiplication that works as an oracle in Shor algorithm in the query model.
To be specific, the goal is to find a quantum circuit for
\begin{align}
  U_{\rm Shor}: \ket{0} \ket{x} \ket{0}_w \mapsto \frac{1}{\sqrt{2}} \Big( \ket{0} \ket{x} \ket{0}_w + \ket{1} \ket{\mathfrak{a} x} \ket{0}_w \Big),
\end{align}
where the left-most ket is a single qubit state (data qubit hereafter), the second register is space for encoding integers, $\mathfrak{a} x$ is $x$ multiplied by a constant $\mathfrak{a}$ modulo $N \in \mathbb{Z}$, and $\ket{0}_w$ is a work register consisting of a certain number of qubits.
Efficient implementation of this kind of operation has been studied thoroughly (see, for example, Refs.~\onlinecite{rines18,gidney19} and related materials therein), but the purpose of this work is not to insist on the optimality.
Readers are assumed to be familiar with quantum arithmetics appearing in Shor algorithm.
Understanding of any explicit circuit construction is helpful, but Refs.~\onlinecite{vbe96,beauregard03} could be the most easily accessible ones.

Assuming a circuit for modular multiplication $U_M: \ket{x}\ket{y}_w \mapsto \ket{x}\ket{y+\mathfrak{a} x}_w$ is given, a naive way to achieve the goal is to make use of $C\textrm{-}U_M$, a controlled version of $U_M$ as follows:
\begin{align}\label{cmul-1}
     &\ket{0} \ket{x} \ket{0}_w \nn
     \overset{H}{\;\;\longmapsto\;\;} & \Big( \ket{0} \ket{x} \ket{0}_w + \ket{1} \ket{x} \ket{0}_w \Big)/\sqrt{2} \nn
     \overset{C\textrm{-}U_M}{\;\;\longmapsto\;\;} & \Big( \ket{0} \ket{x} \ket{0}_w + \ket{1} \ket{x} \ket{0 + \{\mathfrak{a}\} [x]}_w \Big)\sqrt{2} \nn
     \overset{C\textrm{-Swap}}{\;\;\longmapsto\;\;} & \Big( \ket{0} \ket{x} \ket{0}_w + \ket{1} \ket{\mathfrak{a} x} \ket{x}_w \Big)\sqrt{2} \nn
     \overset{C\textrm{-}U_M}{\;\;\longmapsto\;\;} & \Big( \ket{0} \ket{x} \ket{0}_w + \ket{1} \ket{\mathfrak{a} x} \ket{x - \{\mathfrak{a}^{-1}\} [\mathfrak{a} x]}_w \Big)\sqrt{2} ,
\end{align}
where $H$ is a Hadamard gate, $C\textrm{-Swap}$ is a controlled swap operation that roughly costs $n\;(=\lceil \log N \rceil)$ Toffoli gates, and multipliers and multiplicands are surrounded by curly and square brackets, respectively.
The above scheme is frequently adopted in the literature, notably by the circuit designs with a small number of qubits\cite{beauregard03,takahashi06,haner17}.

This scheme can be improved such that the role of $C\textrm{-}U_M$ is replaced by $U_M$.
A controlled version of a certain operation is usually more expensive than its uncontrolled counter-part, and thus replacing $C\textrm{-}U_M$ by $U_M$ likely leads to savings in cost.
Note however that the amount of benefit depends on the underlying addition circuit which is beyond the scope of this report.
The idea is to make use of the value 0 as multiplicands as mentioned in Ref.~\onlinecite{rines18}.
The following procedure may help readers understand it.
The coefficient $1/\sqrt{2}$ induced by Hadamard gate is dropped for simplicity.
\begin{align}\label{cmul-2}
     &\ket{0} \ket{x} \ket{0}_w \nn
     \overset{H}{\;\;\longmapsto\;\;} &\ket{0} \ket{x} \ket{0}_w + \ket{1} \ket{x} \ket{0}_w \nn
     \overset{C\textrm{-cp}}{\;\;\longmapsto\;\;} &\ket{0} \ket{x} \ket{0}_w + \ket{1} \ket{x} \ket{x}_w \nn
     \overset{U_M}{\;\;\longmapsto\;\;} &\ket{0} \ket{x+\{\mathfrak{a}-1\}[0]} \ket{0}_w + \ket{1} \ket{x+\{\mathfrak{a}-1\}[x]} \ket{x}_w  \nn
     \overset{C\textrm{-Swap}}{\;\;\longmapsto\;\;} &\ket{0} \ket{x} \ket{0}_w + \ket{1} \ket{x} \ket{\mathfrak{a} x}_w \nn
     \overset{U_M}{\;\;\longmapsto\;\;} &\ket{0} \ket{x-\{\mathfrak{a}^{-1}\}[0]} \ket{0}_w + \ket{1} \ket{x-\{\mathfrak{a}^{-1}\}[\mathfrak{a} x]} \ket{\mathfrak{a}x}_w \nn
     \overset{C\textrm{-Swap}}{\;\;\longmapsto\;\;} &\ket{0} \ket{x} \ket{0}_w + \ket{1} \ket{\mathfrak{a} x} \ket{0}_w ,
\end{align}
where $C\textrm{-cp}$ is a controlled string copy requiring at most $n$ Toffoli gates.
Compared with Eq.\,(\ref{cmul-1}), the above scheme eliminates a need for controlling two $U_M$ operations at the cost of extra $2n$ Toffoli gates.

\begin{figure}[htbp]
    \centering
    \includegraphics[width=0.48\textwidth]{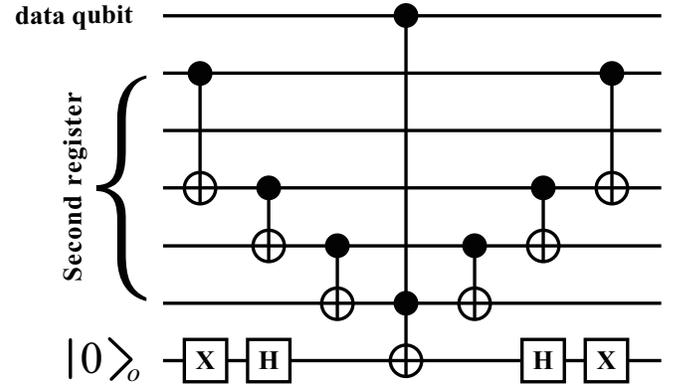}
    \caption{Grover-type oracle $U_f:\ket{y} \mapsto f(y)\ket{y}$ for $\mathfrak{s}=10111$, where $\ket{0}_o$ is an oracle qubit that can be chosen from any idle qubit in the work register.}
    \label{fig:oracle}
\end{figure}

Measurement-based uncomputation can further modify the scheme.
The first five lines in Eq.\,(\ref{cmul-2}) are applied in the same way.
Beginning with $\ket{0}\ket{x}\ket{0}_w + \ket{1}\ket{x}\ket{\mathfrak{a} x}$, we apply $U_M$ to get
\begin{align}\label{cmul-3a}
  &\ket{0}\ket{x}\ket{0}_w + \ket{1}\ket{x}\ket{\mathfrak{a} x} \nn
  \overset{U_M}{\;\longmapsto\;} &
  \ket{0} \ket{x \!-\! \{\mathfrak{a}^{-\!1} \!-\! 1\}[0]} \ket{0}_w
  \!+\!
  \ket{1} \ket{x \!-\! \{\mathfrak{a}^{-\!1} \!-\! 1\}[\mathfrak{a}x]} \ket{\mathfrak{a} x}_w  \nn
  =&\ket{0}\ket{x}\ket{0}_w + \ket{1}\ket{\mathfrak{a} x}\ket{\mathfrak{a} x}_w .
\end{align}
The task is to transform $\ket{\mathfrak{a}x}_w$ into $\ket{0}_w$ with as small number of non-Clifford gates involved as possible.
Phase kick-back technique with an appropriate Deutsch-Jozsa\cite{DJ92} or Grover\cite{grover97} type oracle can be exploited as follows.
Applying Walsh-Hadamard transformation $W\!H$ on the work register of the last expression in Eq.\,(\ref{cmul-3a}) leads to
\begin{align}\label{cmul-3b}
  &\ket{0}\ket{x}\ket{0}_w + \ket{1}\ket{\mathfrak{a} x}\ket{\mathfrak{a} x}_w \nn
  \overset{W\!H}{\;\longmapsto\;} &
  \sum_{s} e^{\pi i ({\vec{s}} \cdot {\vv{0}})} \ket{0} \ket{x} \ket{s}_w
  +
  \sum_{s} e^{\pi i ({\vec{s}} \cdot {\vv{ \mathfrak{a} x} })} \ket{1} \ket{\mathfrak{a}x} \ket{s}_w,
\end{align}
where the symbols with an overhead arrow should be read as vectors in $\mathbb{Z}_2^n$ and dot product is inner product modulo 2.
Normalization constants are still omitted from the expression.
Now measurement in computational basis is carried out on work register giving rise to
\begin{align}\label{cmul-3c}
  \ket{0} \ket{x} \ket{ \mathfrak{s}}_w
  +
  e^{\pi i ({ \vec{\mathfrak{s}}} \cdot \vv{ \mathfrak{a} x} )} \ket{1} \ket{\mathfrak{a}x} \ket{\mathfrak{s}}_w,
\end{align}
which is essentially equivalent to $\ket{0} \ket{x} \ket{0}_w + e^{\pi i ({ \vec{\mathfrak{s}}} \cdot \vv{ \mathfrak{a} x} )} \ket{1} \ket{\mathfrak{a}x} \ket{0}_w$ as the work register is no longer entangled with other qubits.
To get rid of the remaining phase, observe that it is $\pm 1$ depending on $\vec{\mathfrak{s}} \cdot \vv{\mathfrak{a} x}$.
Let $\mathcal{S}=\{\mathfrak{s}_\alpha, \mathfrak{s}_\beta,...\}$ be an ordered set of nonzero digits in the measured string $\mathfrak{s}$, $f(y)=e^{\pi i \vec{ \mathfrak{s}} \cdot \vec{y}}$, and $\mathcal{Y}=\{y_\alpha, y_\beta,...\}$ be an ordered set of digits to be multiplied by $\mathfrak{s}_i \in \mathcal{S}$ upon $\vec{ \mathfrak{s}} \cdot \vec{y}$.
Inspection tells us that
\begin{align}\label{cmul-3d}
   f(y)=
      \left\{
        \begin{array}{ll}
          +1, & \hbox{if $\sum_\alpha y_\alpha \equiv 0$ mod 2;} \\
          -1, & \hbox{if $\sum_\alpha y_\alpha \equiv 1$ mod 2.}
        \end{array}
      \right.
\end{align}
%
It is immediately noticeable that if there exists a quantum oracle $U_f:\ket{y} \mapsto f(y)\ket{y}$, we would have
\begin{align}\label{cmul-3e}
\displaybreak
  &\ket{0} \ket{x} \ket{0}_w + e^{\pi i ({ \vec{\mathfrak{s}}} \cdot \vv{ \mathfrak{a} x} )} \ket{1} \ket{\mathfrak{a}x} \ket{0}_w \nn 
  \overset{U_f}{\;\longmapsto\;} &
  \ket{0} \ket{x} \ket{0}_w + \ket{1} \ket{\mathfrak{a}x} \ket{0}_w.
\end{align}
An oracle can be constructed as follows.
Let us denote positions of nonzero digits in the string $\mathfrak{s}$ by $\alpha,\beta,\gamma,\delta,...,\psi,\omega$.
Beginning with the position $\alpha$ in the second register, apply a controlled-NOT (CNOT) gate with the qubit at $\alpha$ being a control and the qubit at $\beta$ being a target, i.e., CNOT$_{\alpha\beta}$.
Subsequent gates CNOT$_{\beta\gamma}$, CNOT$_{\gamma\delta}$,..., CNOT$_{\psi\omega}$ are also applied in that order.
Applying a Toffoli gate with the data qubit and the qubit at $\omega$ position in the second register being controls and an oracle qubit in $\ket{-}=(\ket{0} + \ket{1})/\sqrt{2}$ state being a target achieves the desired sign change.
Here the data qubit is put on the control position to prevent the sign change in the first term in Eq.\,(\ref{cmul-3e}).
Applied CNOT gates are re-applied in reverse order to recover the string $x$ and $\mathfrak{a}x$, completing the procedure.
Figure\;\ref{fig:oracle} illustrates the procedure assuming the measured string $\mathfrak{s}$ is 10111.

We conclude the report with a complexity analysis.
Compared with Eq.\,(\ref{cmul-1}), the controlled modular multiplication with measurement-based uncomputation replaces two $C\textrm{-}U_M$s by $U_M$s at the cost of $n$ Toffoli gates.
In sum, it will likely save $O(n)$ Toffoli gates, but the advantage is not dramatic as most $U_M$ implementations involve at least $O(n^2)$ Toffoli gates.
Although the specific application examined in this report is not able to fully utilize the measurement-based uncomputation, we believe this direction is worth further investigating.



\bibliographystyle{apsrev4-1}
\bibliography{reference.bib}



\end{document}